\documentclass[12pt]{article}
\usepackage{setspace}
\usepackage[margin=1.1in]{geometry}
\usepackage{mathrsfs}
\usepackage{amsmath}
\usepackage{amssymb}
\usepackage{graphicx} 
\usepackage{deluxetable}
\usepackage[super]{natbib}
\usepackage{float}
\usepackage{bm}
\usepackage{color}
\def\gsim{\;\rlap{\lower 2.5pt \hbox{$\sim$}}\raise 1.5pt\hbox{$>$}\;}
\def\lsim{\;\rlap{\lower 2.5pt  \hbox{$\sim$}}\raise 1.5pt\hbox{$<$}\;}

\def\msun{M_{\odot}}
\def\lsun{L_{\odot}}
\def\mbh{M_{\bullet}}
\def\mh{M_{\rm halo}}
\def\kms{\rm km\,s^{-1}}
\def\ergs{\rm erg\,s^{-1}}
\def\cmc{\rm cm^{-3}}
\def\taugg{\tau_{\gamma \gamma}}
\def\fkin{f_{\rm kin}}
\def\ent{\epsilon_{\rm nt}}
\def\eg{E_{\gamma}}
\def\ep{E_{\rm p}}
\def\lbol{L_{\rm bol}}
\def\ledd{L_{\rm Edd}}
\def\lg{L_{\gamma}}
\def\lgf{L_{\gamma,40}}
\def\lgmax{L_{\gamma,\rm max}}
\def\lgmin{L_{\gamma,\rm min}}
\def\lrad{L_{\rm rad}}

\def\lradtot{L_{\rm rad}^{\rm tot}}
\def\lradtotf{L_{\rm rad,40}^{\rm tot}}
\def\lradcore{L_{\rm rad}^{\rm core}}
\def\lradcoref{L_{\rm rad,40}^{\rm core}}
\def\zmax{z_{\rm max}}
\def\zmin{z_{\rm min}}
\def\fg{F_{\gamma}}
\def\frad{F_{\rm rad}}
\def\fradtot{F_{\rm rad}^{\rm tot}}
\def\fradcore{F_{\rm rad}^{\rm core}}
\def\rhor{\rho_{\rm rad}}
\def\rhogm{\rho_{\gamma}}

\def\tsal{t_{\rm Sal}}
\def\eth{E_{\rm th}}
\def\gp{\Gamma_{\rm p}}
\def\np{n_{\rm p}}
\def\emax{E_{\rm max}}
\def\emin{E_{\rm min}}
\def\lin{L_{\rm in}}
\def\mpt{m_{\rm p}}
\def\tacc{t_{\rm acc}}
\def\tdyn{t_{\rm dyn}}
\def\tpp{t_{\rm pp}}
\def\vs{v_{\rm s}}
\def\rs{R_{\rm s}}
\def\ksib{\xi_{\rm B}}

\def\mtot{M_{\rm tot}}
\def\ms{M_{\rm s}}
\def\pt{P_{\rm T}}
\def\lff{L_{\rm ff}}
\def\lic{L_{\rm IC}}
\def\lsyn{L_{\rm syn}}
\def\lp{L_{\rm p}}
\def\rhog{\rho_{\rm g}}
\def\ng{n_{\rm g}}
\def\fd{f_{\rm d}}
\def\epi{E_{\pi}}
\def\mpi{m_{\pi}}
\def\qpi{q_{\pi}}
\def\sigpp{\sigma_{\rm pp}}
\def\kpi{\kappa_{\rm pp}}

\def\ekin{E_{\rm kin}}
\def\zref{z_{\rm ref}}
\def\rdisk{R_{\rm disk}}
\def\rvir{R_{\rm vir}}
\def\om{\Omega_{\rm M}}
\def\olam{\Omega_{\Lambda}}

\newcounter{firstbib}
\begin{document}

\Large
\centerline{\bf Quasar-driven Outflows Account for the Missing}
\centerline{\bf Extragalactic $\gamma$-Ray Background}

\normalsize
\medskip
\centerline{Xiawei Wang and Abraham Loeb}

\medskip

\centerline{\it Department of Astronomy, Harvard University}
\centerline{\it 60 Garden Street, Cambridge, MA 02138, USA}

\vskip 0.2in
\hrule
\vskip 0.2in

{\bf 
The origin of the extragalactic $\gamma$-ray background permeating throughout the Universe remains a mystery forty years after its discovery\cite{kraushaar1972}.
The extrapolated population of blazars can account for only half of the background radiation at the energy range of $\sim$0.1-10 GeV\cite{ackermann2015}$^{,}$\cite{ajello2015}.
Here we show that quasar-driven outflows generate relativistic protons that produce the missing component of the extragalactic $\gamma$-ray background and naturally match its spectral fingerprint, with a generic break above $\sim$ 1 GeV.
The associated $\gamma$-ray sources are too faint to be detected individually, explaining why they had not been identified so far.
However, future radio observations may image their shock fronts directly.
Our best fit to the \textit{Fermi}-LAT observations of extragalactic $\gamma$-ray background spectrum provides constraints on the outflow parameters that agree with observations of these outflows\cite{fabian2012}$^{-}$\cite{tombesi2015} and theoretical predictions\cite{zubovas2012}$^{,}$\cite{king2015}.
}

The components of the extragalactic $\gamma$-ray background (EGB) has been a puzzle since its discovery four decades ago\cite{fornasa2015}.
Recently, the Large Area Telescope (LAT) on \textit{Fermi} provided fifty-month measurement of the integrated emission from $\gamma$-ray sources, with photon energy extending from 0.1 to 820 GeV\cite{ackermann2015}.
The latest analysis of \textit{Fermi}-LAT data implies that both resolved and unresolved blazars account for $\sim 50^{+12}_{-11}\%$ of the EGB at energy range of $0.1-10$ GeV, leaving the origin of the remaining component in question\cite{ajello2015}.

Active galactic nuclei (AGN) are observed to exhibit strong outflows with velocities of $\sim 0.1\, c$, as manifested by broad absorption lines\cite{arav2013}$^{,}$\cite{tombesi2015}.
The ratio between the input kinetic luminosity of the outflows $\lin$ and the bolometric luminosity of quasars $\lbol$, $\fkin$, is observationally inferred to be $\fkin\sim 1-5\%$ \cite{fabian2012}$^{-}$\cite{tombesi2015}.
The shock wave produced by the interaction of a quasar-driven outflow with the surrounding interstellar medium is expected to accelerate protons to relativistic energies, similarly to the shocks surrounding supernova (SN) remnants where observations of $\gamma$-ray emission due to decay of neutral pion ($\pi^0$) indicate relativistic proton-proton (\textit{pp}) collisions via \textit{pp}$\rightarrow\pi^0\rightarrow2\gamma$\cite{ackermann2013}.
Here, we calculate the analogous $\gamma$-ray emission from quasar-driven outflows.

The energy distribution of accelerated protons per unit volume can be written as
$N(\ep)=N_0\,\ep^{-\gp}$, where $\ep$ is the proton energy, and $N_0$ is a normalization constant, and the power-law idex $\gp\sim 2-3$, based on theoretical models \cite{caprioli2012} and observations of shocks around SN remnants\cite{ackermann2013}$^{,}$\cite{ackermann2015b}.
We adopt a fiducial value of $\gp\sim 2.7$ and show that our results are not very sensitive to variations around it (see \textit{Extended Data} for details).
$N_0$ can be constrained by the total energy condition $\ent\eth=\int_{\emin}^{\emax}\, N(\ep)\ep\, d\ep$, where $\eth$ is the thermal energy density of the shocked particles and $\ent\sim10\%$ is the fraction of the shock kinetic energy converted to accelerate protons\cite{caprioli2014}. 
The minimum energy of the accelerated protons, $\emin$ is set to be the order of proton rest energy $\mpt c^2$ and their maximum energy, $\emax$, is obtained by equating the acceleration time of protons, $\tacc$, to either time scale of \textit{pp} collision, $\tpp\sim(\np\sigpp c)^{-1}\approx 10^8\,n_{\rm p,0}\,\sigma_{\rm pp,-26}\,\rm yrs$, or the dynamical time scale of the outflow shock $\tdyn\sim\rs/\vs\approx 10^6\,R_{\rm s,kpc}\,v_{\rm s,3}^{-1}\,\rm yrs$.
For typical values of the outflows parameters, $\emax$ is on the order of $10^6\,\rm GeV$.
The $e$-folding time to accelerate protons to relativistic energies is $\tacc\sim\ep c/eB\vs^2\approx 300\,E_{\rm p, TeV}\,B_{-6}^{-1}\,v_{\rm s,3}^{-2}\,\rm yrs$, where $E_{\rm p,TeV}=(\ep/\rm TeV)$, $e$ is the electron charge and $B=10^6\,B_{-6}\,\rm G$ is the magnetic field strength\cite{blandford1987}.
We assume a fraction of the post shock thermal energy, $\ksib$, is carried by the magnetic field and adopt $\ksib\sim0.1$, in analogy with SN remnants\cite{chevalier1998}, and we have verified that our results are not sensitive to variations around it.
Here, $n_{\rm p,0}=(\np/1\,\cmc)$ is proton number density, $\sigma_{\rm pp,-26}=(\sigpp/10^{-26}\,\rm cm^2)$ is the inelastic cross section of \textit{pp} collision, $R_{\rm s,kpc}=(\rs/1\,\rm kpc)$ and $v_{\rm s,3}=(\vs/10^3\,\kms)$ are the radius and velocity of the outflowing shell, which can be obtained by solving the hydrodynamics of outflows \cite{wang2015} (see \textit{Methods} for details).

The contribution from quasar outflows to the EGB can be estimated by summing the $\gamma$-ray emission over the known quasar population of all bolometric luminosity at all redshifts.
The cumulative specific intensity is given by:
\begin{equation}
I(\eg)=\iint \Phi(\lbol,z)\frac{\overline{L}_{\gamma}\left(\eg^{\prime},\lbol,z\right)}{4\pi D_{\rm L}^2(z)}\exp\left[-\taugg(\eg^{\prime},z)\right]\frac{dV}{dzd\Omega}\,d\log\lbol\,dz\;,
\end{equation}
where $\eg^{\prime}=\eg(1+z)$ is the intrinsic photon energy, $\Phi(\lbol,z)$ is the quasar bolometric luminosity function\cite{hopkins2007}and $D_{\rm L}(z)$ is the luminosity distance to redshift $z$.
$\overline{L}_{\gamma}\left(\eg,\lbol,z\right)=\tsal^{-1}\,\int L_{\gamma}\left(\eg,\tau,\lbol,z\right)\,d\tau$ is the time-averaged $\gamma$-ray luminosity of an individual quasar outflow, where $\tsal\sim4\times10^7$ yrs is the Salpeter time for a radiative efficiency of $10\%$\cite{yu2002}  (see \textit{Methods} for details).
The diffuse extragalactic background light (EBL) associated with the cumulative UV-optical-infrared emission by star-forming galaxies and AGN over the wavelength range of $0.1-10^3\,\mu$m, attenuates high-energy photons via $e^+e^-$ pair production.
The high energy $\gamma$-ray spectrum is therefore attenuated by photon-photon scattering on the EBL, through a factor of $\exp(-\taugg)$, where $\taugg(\eg,z)$ is the EBL optical depth\cite{stecker2007} for photons with energy $\eg$ at redshift of $z$.

Figure 1 shows the cumulative $\gamma$-ray emission from quasar-driven outflows.
We set upper and lower limits on the contribution from radio galaxies to the EGB based on the most recent \textit{Fermi}-LAT catalog (3FGL) \cite{ackermann2015c} and find that radio galaxies can account for $\sim 7\pm4\%$ of the EGB at $\eg\lesssim10$ GeV, roughly $2-5$ times less than previous estimate based on sources identified in the first and second \textit{Fermi}-LAT catalog (1FGL\cite{inoue2011} \& 2FGL\cite{dimauro2014}, see \textit{Methods} for details).
Star-forming galaxies has been evaluated to constitute $\sim 13\pm9\%$ of the EGB\citep{ackermann2012}. 
We show that the contribution from quasar outflows takes up the remaining $\sim 20-40\%$ of the EGB, which dominates over the total of radio galaxies and star-forming galaxies, and can naturally account for the amplitude and spectral shape of the remaining EGB, while at higher energies the EGB is dominated by blazars\cite{ajello2015}.
We have verified that the cumulative contribution from radio galaxies and star-forming galaxies does not match the EGB's spectral shape in that the EGB would be overproduced at $\eg\gtrsim10$ GeV if the sum of radio galaxies and star-forming galaxies makes up the missing component at $\eg\lesssim10$ GeV.
We find that the break in the spectral energy distribution (SED) of quasar outflows at $\lesssim10$ GeV is independent of the parameter choices for the outflow dynamics.
This generic cutoff in the emission spectrum of quasar outflows naturally fits the missing EGB component.

The fraction of the shock kinetic energy used to accelerate protons $\ent$ and the fraction of the quasar's bolometric luminosity that powers the outflow $\fkin$ are free parameters whose product  $\eta=\ent\fkin$ can be constrained by the EGB data.
We search the minimum of
$\chi^2=\sum_{i=1}^{N}\,\left(I_{\rm obs}^i-I_{\rm mod}^i\right)^2/\Delta_i^2$
throughout the parameter space, where $N$ is the number of data points, $\Delta_i$ is the error bar of the $i^{\rm th}$ observed point and $I_{\rm obs}^i$ and $I_{\rm mod}^i$ are the EBG intensity of the observed and expected values, respectively.
We find the best fit value of $\eta = \left(3.98\pm 0.76\right)\times10^{-3}$ at $90\%$ significance.
For $\ent\sim10\%$ as inferred from observations of SN remnants\cite{ackermann2013} and theoretical models\cite{caprioli2012}$^{,}$\cite{caprioli2014}, we deduce a value of $\fkin\sim1-5\%$, which agrees well with observations of outflows\cite{arav2013}$^{-}$\cite{tombesi2015} and theoretical predictions\cite{zubovas2012}$^{,}$\cite{king2015}.

The bright phase of the $\gamma$-ray emission from an individual quasar ends abruptly when the outflow exits from the surrounding galactic disk, as shown in Figure 2, making it difficult to detect afterwards.
Outflows embedded in Milky Way (MW) mass halos propagating to 10 kpc scale are expected to produce GeV $\gamma$-ray emission of $\sim 10^{39}-10^{40}\,\ergs$.
In the local Universe ($z<0.1$), we find that only $\lesssim 0.1\%$ of quasars host $\gamma$-ray bright outflows that are detectable by \textit{Fermi}-LAT at GeV energies.
These outflows are too faint to be detected in $\gamma$-rays individually, explaining why they have not been identified so far.
A possible candidate of galactic outflow relic is the Fermi bubbles at the Galactic center \citep{su2010}, whose $\gamma$-ray emission has been explained by hadronic process similar to our model \citep{crocker2011}$^{,}$\citep{crocker2015}.
Our interpretation can be tested through observations of quasar outflows at other wavelengths.
Radio emission is simultaneously produced via synchrotron from accelerated electrons by the same outflow shocks (see black solid line in Fig.2).
Radio telescopes such as \textit{the Jansky Very Large Array} and \textit{the Square Kilometre Array} provide high sensitivity to detect this emission and confirm the parameters of outflows\cite{wang2015} at redshifts up to $\sim$ 5.
For most AGNs, the radio emission is free of contamination from the central source or scattering of its light by surrounding electrons.
Source stacking \cite{cillis2004} could be performed in the future to find direct evidence for the cumulative $\gamma$-ray signal from multiple outflow-hosting quasars.
\vskip 0.45in
\hrule

\normalsize
\vskip 0.3in
\noindent
{ACKNOWLEDGEMENTS.} 
\\
We thank Douglas Finkbeiner, James Guillochon and Lorenzo Sironi for helpful comments on the manuscript.
We thank Jonathan Grindlay, Vinay Kashyap, Mark Reid, Aneta Siemiginowska and Stephen Portillo for useful discussion.
This work is supported by NSF grant AST-1312034.

\normalsize
\vskip 0.3in
\noindent

\vskip 1in
\begin{figure}[H]
\includegraphics[angle=0,width=\columnwidth]{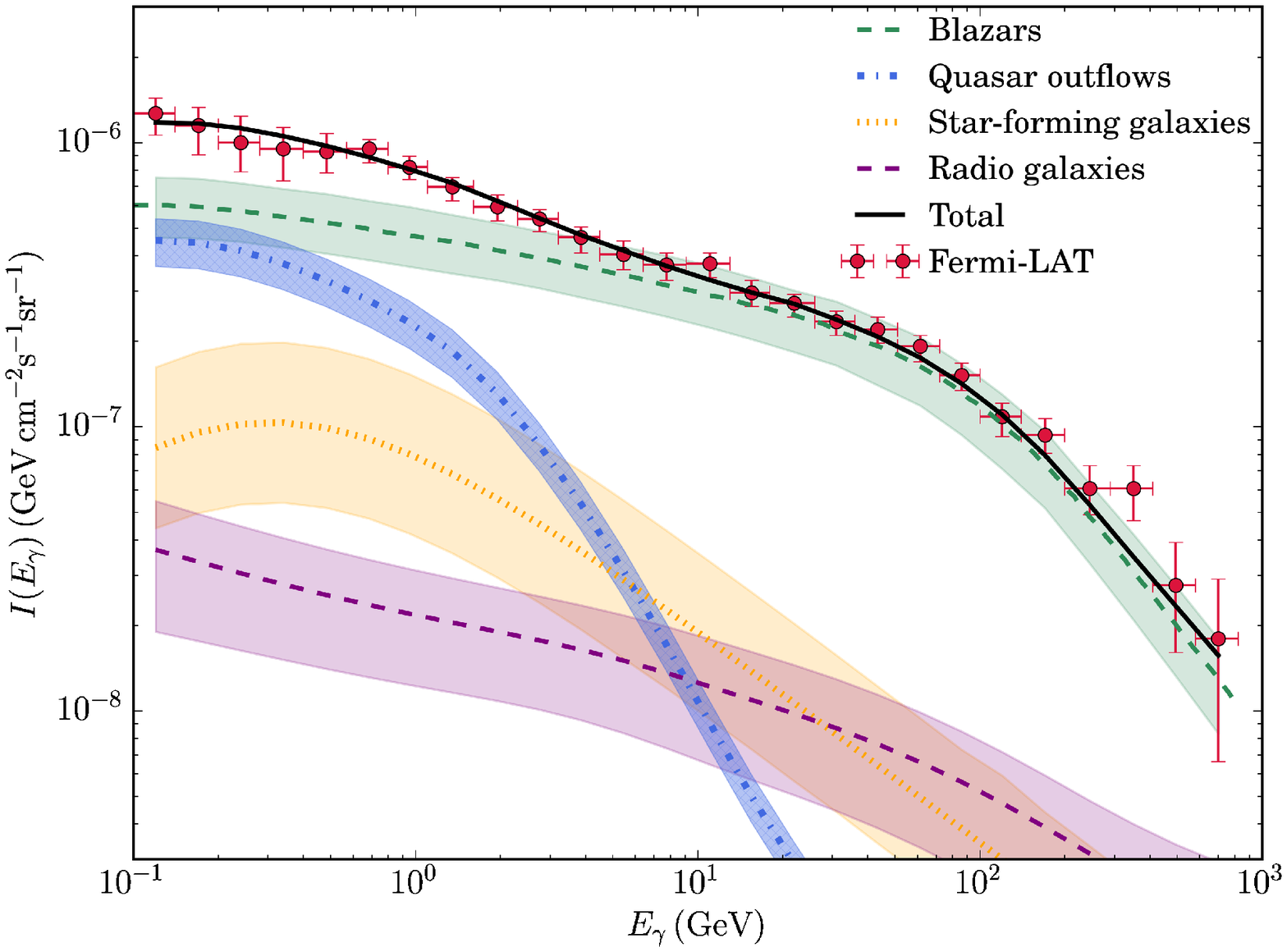}
\caption{
Spectral energy distribution of the integrated $\gamma$-ray background.
\textit{Fermi}-LAT data of the extragalactic $\gamma$-ray background is shown as the red points with error bars taken from Ackermann et al. (2015)\citep{ackermann2015}.
The green dashed line corresponds to the contribution from blazars as estimated by Ajello et al. (2015)\citep{ajello2015}.
The purple dashed line shows the contribution from radio galaxies, following Inoue (2011)\citep{inoue2011} and Di Mauro et al. (2014)\citep{dimauro2014}, derived from the most recent sample in \textit{Fermi}-LAT catalog \citep{ackermann2015c} (Ackermann et al. 2015).
The orange dotted line corresponds to the contribution from star-forming galaxies as estimated by Ackermann et al. (2012)\citep{ackermann2012}, assuming $\gamma$-ray emission spectral shape follows that of the MW. 
The dot-dashed blue line represents the contribution from our quasar-driven outflow model with $\eta_{-3}=3.98$, where $\eta_{-3}=(\eta/10^{-3})$.
The total contribution to EGB from all sources is shown as the solid black line.
The shaded regions indicate the uncertainties of each component.
}
\end{figure}

\vskip 1in
\begin{figure}[H]
\includegraphics[angle=0,width=\columnwidth]{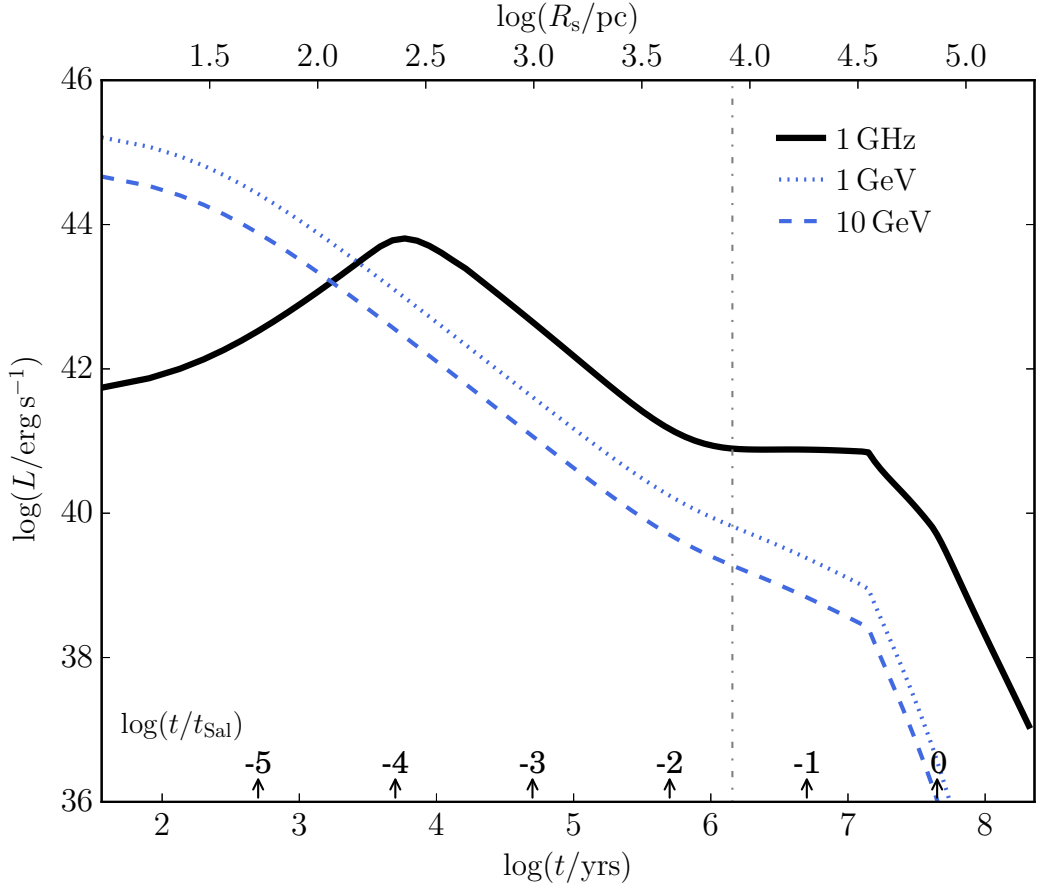}
\caption{
Light curve of $\gamma$-ray emission from AGN-driven outflows and its radio counterpart, for a halo mass $\mh=10^{12}\,\msun$ and redshift $z=0.1$.
The solid black line represents the radio synchrotron emission at 1 GHz from electrons accelerated at the outflow shock front\citep{wang2015}.
The dotted and dashed blue lines show the $\gamma$-ray emission from accelerated protons with photon energies at 1 GeV and 10 GeV, respectively.
The dot-dashed vertical line marks the transition of the outflow from the disk to the halo of its host galaxy.
The radio and $\gamma$-ray luminosity are shown as a function of time, $t$, and outflow shock radius, $\rs$, on the lower and upper horizontal axes, respectively.
Above the lower horizontal axis, we express the time as a fraction of the Salpeter time $\tsal$, indicating roughly the probability of finding a quasar at each time or position.
The vast majority of the quasar outflows are too faint to be detected individually, explaining why their contribution to the EGB had not been recognized.
}
\end{figure}
%

\newpage
\vskip 1in
\Large
\textbf{Methods}

\normalsize
\textbf{Hydrodynamics of quasar-driven outflows.}
Outflows are injected into the ambient medium with an initial velocity of $\sim0.1\,c$.
As they propagate in the ambient medium, a double shock structure forms.
The outer forward shock accelerates the swept-up medium while the inner reverse shock decelerate the wind itself. 
The equations describing outflow hydrodynamics are given by\citep{wang2015}$^{,}$\citep{fgq2012}:
\begin{subequations}
\begin{equation}
\frac{d^2 \rs}{dt^2}=\frac{4\pi\rs^2}{\ms}(\pt-P_0)-\frac{G\mtot}{\rs^2}-\frac{\vs}{\ms}\frac{d\ms}{dt}\, ,
\end{equation}
\begin{equation}
\frac{d\ms}{dt}=4\pi\rhog\rs^2\vs\, ,
\end{equation}
\begin{equation}
\frac{d\pt}{dt}=\frac{\Lambda}{2\pi\rs^3}-5\pt\frac{\vs}{\rs}\, ,
\end{equation}
\begin{equation}
\Lambda=\lin-\lff-\lic-\lsyn-\lp\, ,
\end{equation}
\end{subequations}
where $\ms$ is the swept-up mass of the outflowing shell.
$\mtot$ is the gravitational mass decelerating the expansion of  the shell, including the  mass of dark matter, host galaxy, central black hole (BH) and the self gravity of the shell.
The thermal pressure in the shocked wind $\pt$ declines at a rate determined by the work done on the medium and radiative losses in the shocked wind, described by the heating/cooling luminosity $\Lambda$, composed of energy injection $\lin$, free-free emission $\lff$, synchrotron cooling $\lsyn$, inverse Compton scattering $\lic$ and proton cooling $\lp$.
Hydrostatic equilibrium gives the thermal pressure in the ambient medium $P_0$.
We assume that quasars radiate at Eddington luminosity $\ledd\approx 1.3\times10^{38}(\mbh/\msun)\,\ergs$, providing BH mass $\mbh$ for a given $\lbol$.
The prescription for assigning $\mbh$ to host halo mass $\mh$ follows Guillochon \& Loeb\cite{guillochon2015} (2015) with a fixed bulge to total stellar mass ratio of 0.3. 
We set an upper limit of $\mh$ to be $\sim 10^{13}\msun$, which is the maximum halo mass that allows the propagation of outflows to halo scale as found in numerical results\cite{wang2015}.
We make the assumption of spherical symmetry for the density distribution of the surrounding gas and the mass profile of the galaxy where the outflows are embedded.
The gas density distribution is assumed to be a broken power-law, expressed as:
\begin{eqnarray}
\rhog(R)\propto
\begin{cases}
R^{-\alpha}		& (R<\rdisk) \\
R^{-\beta}		& (\rdisk<R<\rvir) \\
\end{cases}
\end{eqnarray}
where $\alpha$ and $\beta$ are different indices for the disk and halo components and $\rdisk$ and $\rvir$ are the radius of the disk and halo, respectively.
We fix $\alpha=2$ assuming an isothermal sphere for the gas within the disk component and $\beta$ can be self-consistently constraint by halo mass $\mh$, redshift $z$ and disk baryonic fraction $\fd$ (taken to be 0.5 in the calculation).

\textbf{$\gamma$-ray spectrum from quasar-driven outflows.}
We compute the spectral energy distribution (SED) of gamma-ray emission produced by neutral pion ($\pi^0$) decay.
For $\ep\lesssim0.1$ TeV, the $\gamma$-ray luminosity is given by\citep{aharonian2000}:
\begin{equation}
L(\eg)=2V\eg^2\,\int_{\emin}^{\infty}\,\frac{\qpi(\epi)}
{\sqrt{\epi^2-\mpi^2 c^4}}\,d\epi\; ,
\end{equation}
where $\emin=\eg+\mpi^2 c^4/4\eg$, $\mpi$ and $\epi$ are the mass and energy of $\pi^0$ and $V$ is the volume of the outflow.
$\qpi(\epi)$ is the emissivity of $\pi^0$, given by\citep{aharonian2000}:
\begin{equation}
\qpi(\epi)=\frac{c\ng}{\kpi}\sigpp(x) N(x)\; ,
\end{equation}
where $x=\mpt c^2+\epi/\kpi$, $\kpi\sim 17\%$ is the fraction of the relativistic proton energy that goes to neutral pions in each interaction, $N(x)$ is the energy distribution of accelerated protons and $\ng=\rhog/\mpt$ is the number density of the ambient medium.
The inelastic cross section of \textit{pp} collision $\sigpp$ is approximated by\citep{aharonian2000}:
\begin{equation}
\sigpp(\ep)\approx 30\left[0.95+0.06\ln(\ekin/\rm GeV)\right]\,\rm mb\; ,
\end{equation}
for $\ekin\ge 1$ GeV, and $\sigpp=0$ is assumed at lower energies, where $\ekin=\ep-\mpt c^2$ is the kinetic energy of protons.
This implies that the $\gamma$-ray emission is produced by relativistic protons with energy $\gtrsim 2$ GeV.
We have verified that the variation in results adopting other approximations of $\sigpp$ is negligible\cite{kelner2006}.
We estimate that the timescale of Coulomb collisions\cite{sturner1997} is $\sim 10$ times longer than $\tpp$, meaning that \textit{pp} collisions is the dominant proton cooling process.
The $\gamma$-ray SED of an individual quasar outflow for different power-law indices of accelerated protons $\Gamma_{\rm p}$ is shown in Fig.2 in \textit{Extended Data}.
For a quasar with halo mass $\sim 10^{12}\,\msun$ at redshift $z\sim 0.1$, the expected GeV $\gamma$-ray luminosity is $\sim 10^{39}-10^{40}\,\ergs$, 
which falls off the detection limit of \textit{Fermi}-LAT by $\sim 2-3$ orders of magnitude.

\textbf{Integrated $\gamma$-ray background.}
The bolometric luminosity function of quasars is given by\citep{hopkins2007}:
\begin{equation}
\Phi(\lbol,z)=\frac{\Phi_{\star}}{(\lbol/L_{\star})^{\gamma_1}+(\lbol/L_{\star})^{\gamma_2}};,
\end{equation}
where $\lbol$ is the bolometric luminosity, $L_{\star}$ varies with redshift, described by 
$\log L_{\star}=(\log L_{\star})_0+k_{L,1}\xi+k_{L,2}\xi^2+k_{L,3}\xi^3$, $\xi=\log[(1+z)/(1+\zref)]$, $\zref=2$ and $k_{L,1}$, $k_{L,2}$ and $k_{L,3}$ are free parameters.
We adopt parameter values of the pure luminosity evolution model, where $\log(\Phi_{\star}/\rm Mpc^{-3})=-4.733$, $(\log (L_{\star}/\lsun))_0=12.965$, $\lsun=3.9\times10^{33}\,\ergs$, $k_{L,1}=0.749$, $k_{L,2}=-8.03$, $k_{L,3}=-4.40$, $\gamma_1=0.517$ and $\gamma_2=2.096$.
We integrate the $\gamma$-ray emission over bolometric luminosity range $10^{42}-10^{48}\,\ergs$ and redshift range $0-5$.
The comoving volume per unit redshift is given by\cite{carroll1992}:
\begin{equation}
\frac{dV}{dzd\Omega}=D_{\rm H}\frac{D_{\rm L}^2(z)}{(1+z)^2 E(z)}\, ,
\end{equation}
where $D_{\rm H}= c/H_0$ and $E(z)=\sqrt{\om(1+z)^3+\olam}$.
We adopt $H_0=70\,\kms\,\rm Mpc^{-1}$, $\om=0.30$ and $\olam=0.7$.

\textbf{Constraints on radio galaxies' contribution to the EGB.}
We estimate the contribution to the EGB by radio galaxies (RGs) using samples identified in the most recent \textit{Fermi}-LAT catalog, 3FGL\citep{ackermann2015c}.
Compared with previous \textit{Fermi}-LAT catalogues, PKS 0943-76 has been removed due to misassociation \citep{ackermann2015c}.
The association of Fornax A (NGC 1316) has not been confirmed by 3FGL \citep{ackermann2015c}.
Newly identified FRI (Fanaroff-Riley type I) sources include 4C+39.12 and 3C 264, and FRII sources include 3C 303, 3C 286 and 3C 275.1.
Consequently, 19 objects constitute our RG sample and their parameters are summarized in Table 1.
We note that some FRI sources such as IC 310, PKS 0625-35 and NGC 1275 show blazar-like variabilities, which could lead to debatable source classification with BL Lac objects.
TXS 0348+013, 3C 207, 3C 275.1, 3C 286 and 3C 380 are classified as steep-spectrum radio quasars (SSRQs) and thus are non-standard FRIIs \cite{dimauro2014}.
However, FRI/BLL and SSRQ sources are also included in sample selection of Inoue (2011)\citep{inoue2011} and Di Mauro et al. (2014)\citep{dimauro2014}.
Therefore, we keep them in our source selection in consistency with previous analysis and we have verified that removal of them lead to negligible change in radio-$\gamma$-ray correlation as discussed later.

Previously, the contribution of RGs to the EGB has been evaluated based on the $\gamma$-ray luminosity function of RGs, which is established from a correlation between $\gamma$-ray and 5 GHz \textit{core-only} radio luminosities of RGs \cite{inoue2011}.
However, the origin of the $\gamma$-ray emission from RGs remains uncertain.
$\gamma$-ray emission could be produced by ultrarelativistic electrons of high density in the radio lobes by scattering soft photons via self-synchrotron Compton or external Compton processes.
Such $\gamma$-ray emission has been resolved and confirmed in the lobes of a nearby FRI RG, Cen A by \textit{Fermi}-LAT \citep{abdo2010b}.
Due to the lack of simultaneous radio and $\gamma$-ray observations of RGs, core variabilities could invalidate this correlation.
In our calculation below, we choose radio data closest in date to $\gamma$-ray observations.
The correlation between the \textit{core-only} radio luminosity and the total $\gamma$-ray luminosity would be distorted if some of the unresolved $\gamma$-ray emission originates outside the core of the corresponding galaxies.
In such a case, the $\gamma$-ray emission from the core would be overestimated, and the radio-$\gamma$-ray correlation would provide an upper limit on the contribution of RGs to the EGB.
The actual contribution would be between this upper limit and the result one gets when correlating the \textit{total} radio and $\gamma$-ray emission of these galaxies.

We recalculate the $\lg$-$\lrad$ correlation for both \textit{core-only} and \textit{total} radio luminosity cases using the most recent samples.
We follow the BCES (bivariate correlated errors and intrinsic scatter) method by Akritas \& Bershady (1996) \citep{akritas1996} to fit regression parameters and uncertainties.
Using the BCES($\lg\mid\lrad$) slope estimator,  we find that the best fit $\lg$-$\lradtot$ and $\fg$-$\fradtot$ correlation can be expressed as:
\begin{subequations}
\begin{equation}
\log(\lgf)=(0.972\pm0.087)\log(\lradtotf)+(1.944\pm0.233)\,,
\end{equation}
\begin{equation}
\log(\fg)=(0.682\pm0.185)\log(\fradtot)+(-11.330\pm0.141)\;,
\end{equation}
\end{subequations}
where $\lgf$ and $\lradtotf$ are $\lg$ and total radio luminosity $\lradtot$ in units of $10^{40}\,\ergs$.
Similarly, the best fit $\lg$-$\lradcore$ and $\fg$-$\fradcore$ are given by:
\begin{subequations}
\begin{equation}
\log(\lgf)=(0.934\pm0.073)\log(\lradcoref)+(2.582\pm0.103)\,,
\end{equation}
\begin{equation}
\log(\fg)=(0.790\pm0.183)\log(\fradcore)+(-10.910\pm0.106)\;,
\end{equation}
\end{subequations}
where $\lradcore$ is the core-only radio luminosity in units of $10^{40}\,\ergs$.
$\gamma$-ray-radio correlations based on 1FGL\citep{inoue2011} and 2FGL\citep{dimauro2014} samples are given by:
\begin{equation}
\log(\lgf)=(1.16\pm0.02)\log(\lradcoref)+(2.5\pm1.41)\,,
\end{equation}
\begin{equation}
\log(\lgf)=(1.008\pm0.025)\log(\lradcoref)+(2.32\pm1.98)\,.
\end{equation}
We compare our fitted $\lg$-$\lradtot$ and $\lg$-$\lradcore$ correlation with previous results, shown in \textit{Extended Data} Fig.3 and Fig.4, respectively. 
We calculate the corresponding Spearman coefficients and partial correlation coefficient of $\lg$ and $\lrad$, $\fg$ and $\frad$ and the corresponding $p$-values, summarized in Table 2.

Following Inoue (2011)\citep{inoue2011} and Di Mauro et al. (2014) \citep{dimauro2014}, we calculate RG's contribution to the EGB using our updated $\gamma$-ray-radio correlation.
The $\gamma$-ray luminosity function (GLF) can be obtained by:
\begin{equation}
\rhogm=\kappa\rhor\frac{d\log\lrad}{d\log\lg}\;,
\end{equation}
where $\rhor$ is the radio luminosity function (RLF) of RGs, and $\kappa$ is the fraction of $\gamma$-ray loud RGs, constraint by source-count distribution as discussed later in the text.
For the \textit{total}-radio-$\gamma$-ray luminosity correlation, we adopt the total RLF and corresponding parameters given by model C of Willott et al. (2001) \citep{willott2001} and convert it to the cosmological constants in this work.
For the \textit{core-only} radio luminosity correlation, we convert the total RLF to core RLF, following the method proposed by Di Mauro et al. (2014) \citep{dimauro2014}, according to the core-total radio luminosity correlation of RGs \citep{lara2004}:
\begin{equation}
\log L_{\rm rad,core}^{5\,\rm GHz} = (0.77\pm0.08)\log L_{\rm rad,tot}^{1.4\,\rm GHz}+(4.2\pm2.1)\;,
\end{equation}
where core radio luminosity at 5 GHz $L_{\rm rad,core}^{5\,\rm GHz}$ and total radio luminosity at 1.4 GHz $L_{\rm rad,tot}^{1.4\,\rm GHz}$ are in units of W Hz$^{-1}$.
We adopt a radio spectral index $\alpha_{\rm r}=0.8$ for conversion of radio luminosities at different frequencies in our calculation \citep{willott2001}.

The intrinsic $\gamma$-ray photon flux per unit energy is obtained by:
\begin{equation}
\frac{dS_{\gamma}}{d\eg}(\eg,\lg,z,\Gamma)=\frac{2-\Gamma}{E_1^2}\left(\frac{\eg}{E_1}\right)^{-\Gamma}
\left[\left(\frac{E_2}{E_1}\right)^{2-\Gamma}-1\right]^{-1}
\frac{\lg(1+z)^{2-\Gamma}}{4\pi D_{\rm L}^{2}(z)}\;,
\end{equation}
where $\Gamma$ is the $\gamma$-ray photon index.
Therefore, we obtain the integrated $\gamma$-ray SED from RGs, expressed as:
\begin{equation}
\begin{split}
I(\eg) = & \eg^2\int_{\Gamma_{\rm min}}^{\Gamma_{\rm max}}\frac{dN_{\Gamma}}{d\Gamma}d\Gamma
\int_{\zmin}^{\zmax}\frac{dV}{dzd\Omega}\,dz
\int_{\lgmin}^{\lgmax}d\log\lg\,\rhogm(\lg,z)\,\\
& \times\frac{dS_{\gamma}}{d\eg}\left(\eg^{\prime},z,\lg,\Gamma\right)
\exp\left[-\taugg(\eg^{\prime},z)\right] \{1-\omega\left[S_\gamma(\lg,z)\right]\}\,,
\end{split}
\end{equation}
where $dN_{\Gamma}/d\Gamma$ is the distribution of $\gamma$-ray photon index $\Gamma$, which is assumed to be Gaussian in an analogy to blazars, with an average value of 2.25 and a scatter of 0.28 based on our RG sample.
$\omega(S_\gamma)$ is the detection efficiency of \textit{Fermi}-LAT at a photon flux of $S_\gamma$.
However, $\omega(S_\gamma)$ is not given in 3FGL, so we adopt the derived detection efficiency for detection threshold TS $>25$ and $|b|<10^{\circ}$ derived for 2FGL\citep{dimauro2014}.
We adopt $\Gamma_{\rm min}=1.0$, $\Gamma_{\rm max}=5.0$, $\zmin=0.0$, $\zmax=5.0$, $\lgmin=10^{38}\,\ergs$ and $\lgmax=10^{50}\,\ergs$ in our calculation.

The expected cumulative flux distribution can be obtained by:
\begin{equation}
N_{\rm exp}(>S_{\gamma}) = 4\pi\int_{\Gamma_{\rm min}}^{\Gamma_{\rm max}}
\frac{dN_{\Gamma}}{d\Gamma}\,d\Gamma \int_{\zmin}^{\zmax}\frac{dV}{dzd\Omega}\,dz
\int_{\lg(S_\gamma,z)}^{\lgmax} \rhogm(\lg,z)\,d\log\lg\;,
\end{equation}
where $S_{\gamma}$ is the photon flux above 0.1 GeV and $\lg(S_\gamma,z)$ is the corresponding $\gamma$-ray luminosity at a redshift of $z$.
The observed source-count distribution of our sample is given by \cite{abdo2010c}:
\begin{equation}
N_{\rm obs}(>S_{\gamma})=\sum_{i=1}^{N(>S_{\gamma,i})}\,\frac{1}{\omega(S_{\gamma,i})}\;,
\end{equation}
where we sum up all RG sources with photon flux $S_{\gamma,i}>S_{\gamma}$.
$\kappa$ can be constraint by normalizing $N_{\rm exp}$ to $N_{\rm obs}$.
We find the best fit at 1$\sigma$ significance is $\kappa=0.081\pm0.008$ by using \textit{total}-radio-$\gamma$-ray luminosity correlation (Eq.(9)), and $\kappa=2.32\pm0.15$ by using \textit{core}-radio-$\gamma$-ray luminosity correlation (Eq.(10), (14)).
This indicates that the \textit{core-only} radio-$\gamma$-ray correlation overproduces $\gamma$-ray loud RGs constraint by the observed source-count distribution.
In this case, we fix $\kappa=1$ in our calculation following Di Mauro et al. (2014) \citep{dimauro2014}.

We obtain the resulting integrated $\gamma$-ray spectrum for both cases, which set the upper and lower limits of RG's contribution to the EGB.
In our calculation, we adopt the mid-value of this range as RG's contribution and show the full range as uncertainty.

We find that the RGs make up $\sim7\pm4\%$ of the EGB. 
We have verified that if RGs accounts for the rest of the EGB besides blazars and star-forming galaxies at $\eg\lesssim10$ GeV, then the EGB would be overproduced at higher energies.
However, quasar outflow's SED has a generic break at $< 10$ GeV, which naturally account for the missing component of the EGB. 
\vskip 0.45in
\hrule
\vskip 0.15in
%

%

\newpage
\vskip 1in
\Large
\textbf{Extended Data}
\setcounter{figure}{0}
\normalsize
We compare the integrated SED for different power-law indices of accelerated protons $\Gamma_{\rm p}$, shown below.
\begin{figure}[H]
\begin{centering}
\includegraphics[angle=0,width=\columnwidth]{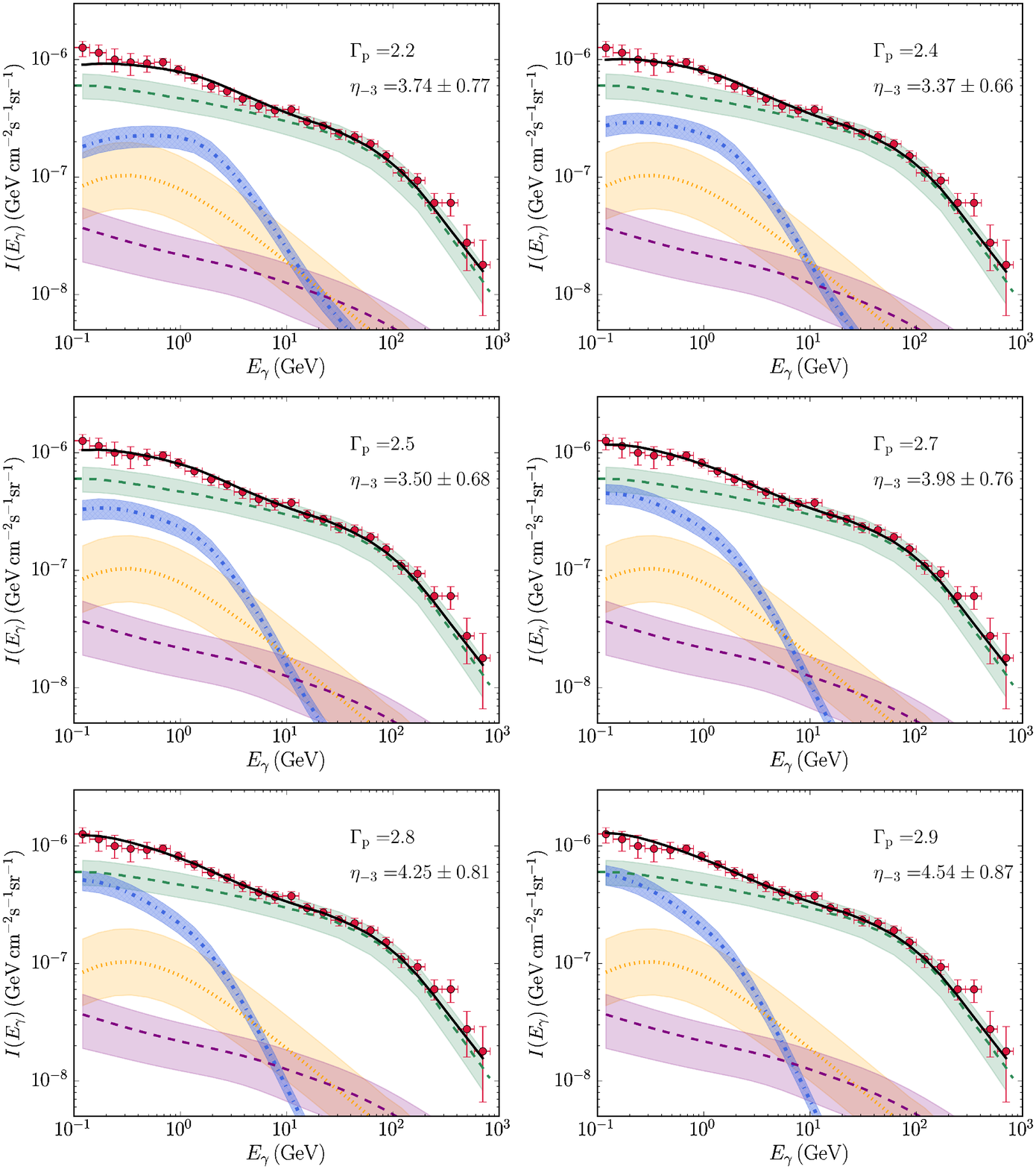}
\caption{
Dependence of the spectral energy distribution of the integrated $\gamma$-ray background on accelerated proton distribution index $\gp$.
Legends are the same as in Fig.1.
We adopt $\gp=$2.2, 2.4, 2.5, 2.7, 2.8, 2.9 and show the best fit $\eta_{-3}$  at 90\% significance correspondingly.
}
\end{centering}
\end{figure}
\begin{figure}[H]
\begin{centering}
\includegraphics[angle=0,width=\columnwidth]{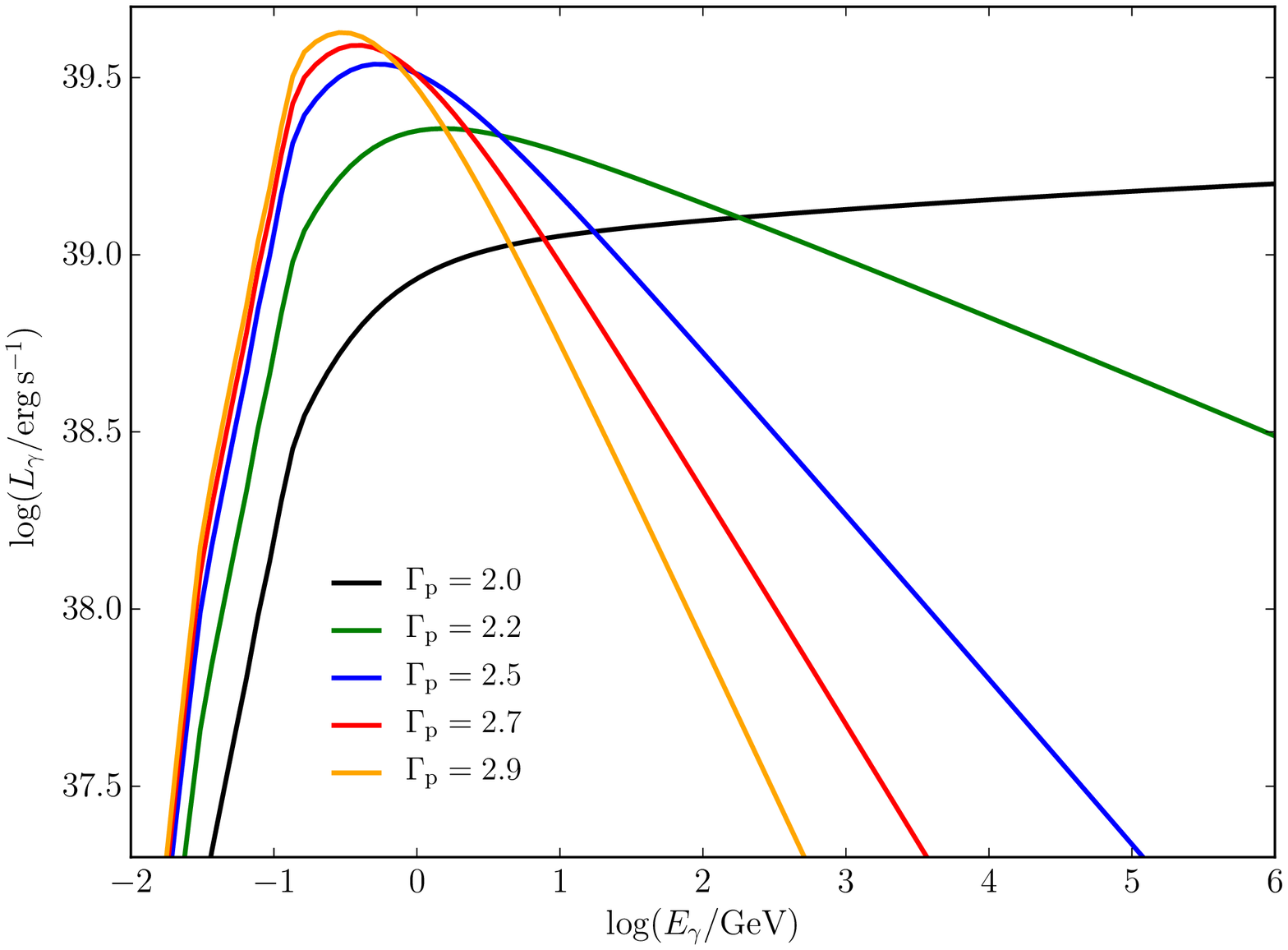}
\caption{
Spectral energy distribution of an individual quasar outflow embedded in a halo of mass $10^{12}\,\msun$ at redshift 0.1.
The black, green, blue, red and orange lines correspond to power-law indices of accelerated protons $\Gamma_{\rm p}=2.0$, 2.2, 2.5, 2.7 and 2.9, respectively.
}
\end{centering}
\end{figure}
\begin{figure}[H]
\begin{centering}
\includegraphics[angle=0,width=\columnwidth]{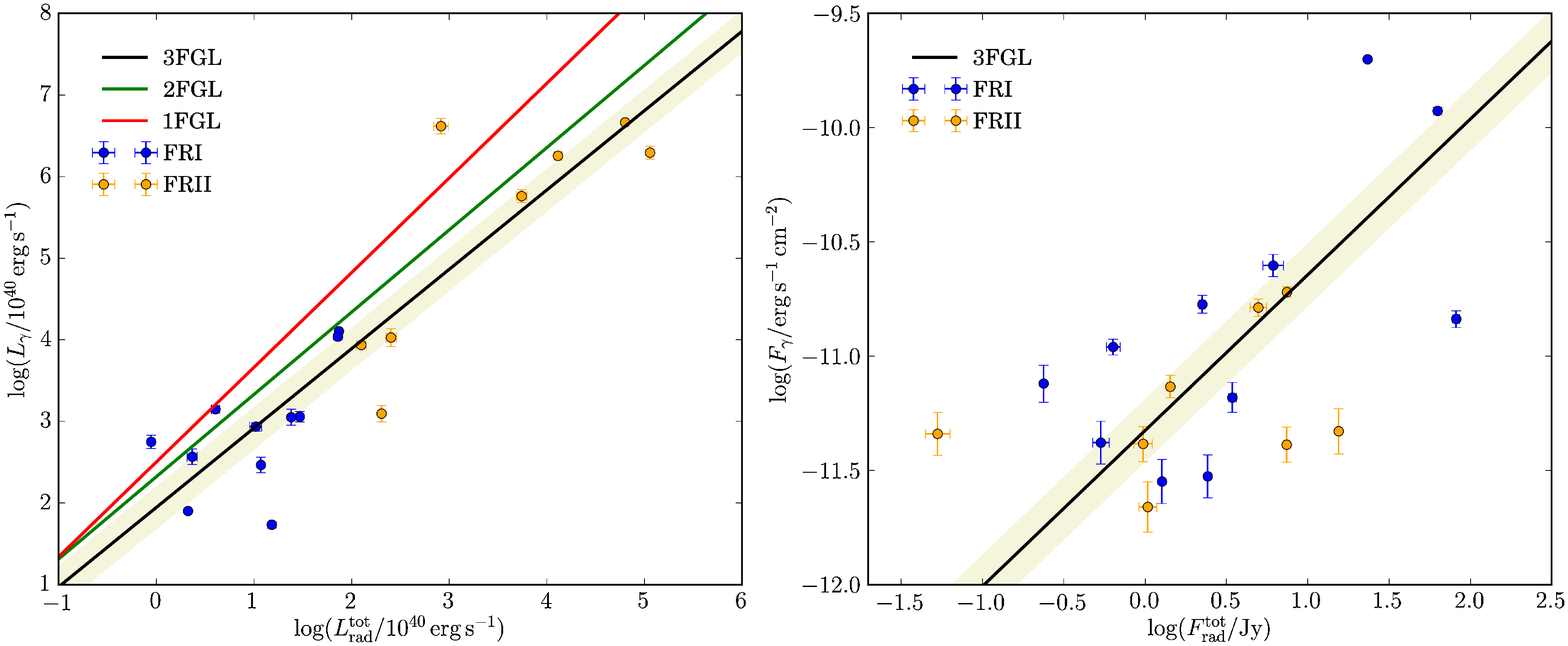}
\caption{Correlation between $\gamma$-ray luminosity/flux above 100 MeV and \textit{total} radio luminosity/flux at 5 GHz of RGs.
The blue and orange points with error bars represent FRI and FRII type RGs selected from \textit{Fermi}-LAT third catalog (3FGL)\citep{ackermann2015c}.
On the left panel, the red and blue lines show the $\lg$-$\lradcore$ correlation fitted by Inoue (2011) \citep{inoue2011} based on 1FGL and Di Mauro et al. (2014) \citep{dimauro2014} based on 2FGL used to evaluate the $\gamma$-ray luminosity function of RGs, respectively, while the black lines represent the fit using samples from the third \textit{Fermi}-LAT catalog (3FGL) \citep{ackermann2015c}, using BCES sampler\citep{akritas1996}.
On the right panel, we show the correlation between radio and $\gamma$-ray fluxes using the same linear regression sampler.
The shaded beige bands indicate 1$\sigma$ uncertainty.
}
\end{centering}
\end{figure}
\begin{figure}[H]
\begin{centering}
\includegraphics[angle=0,width=\columnwidth]{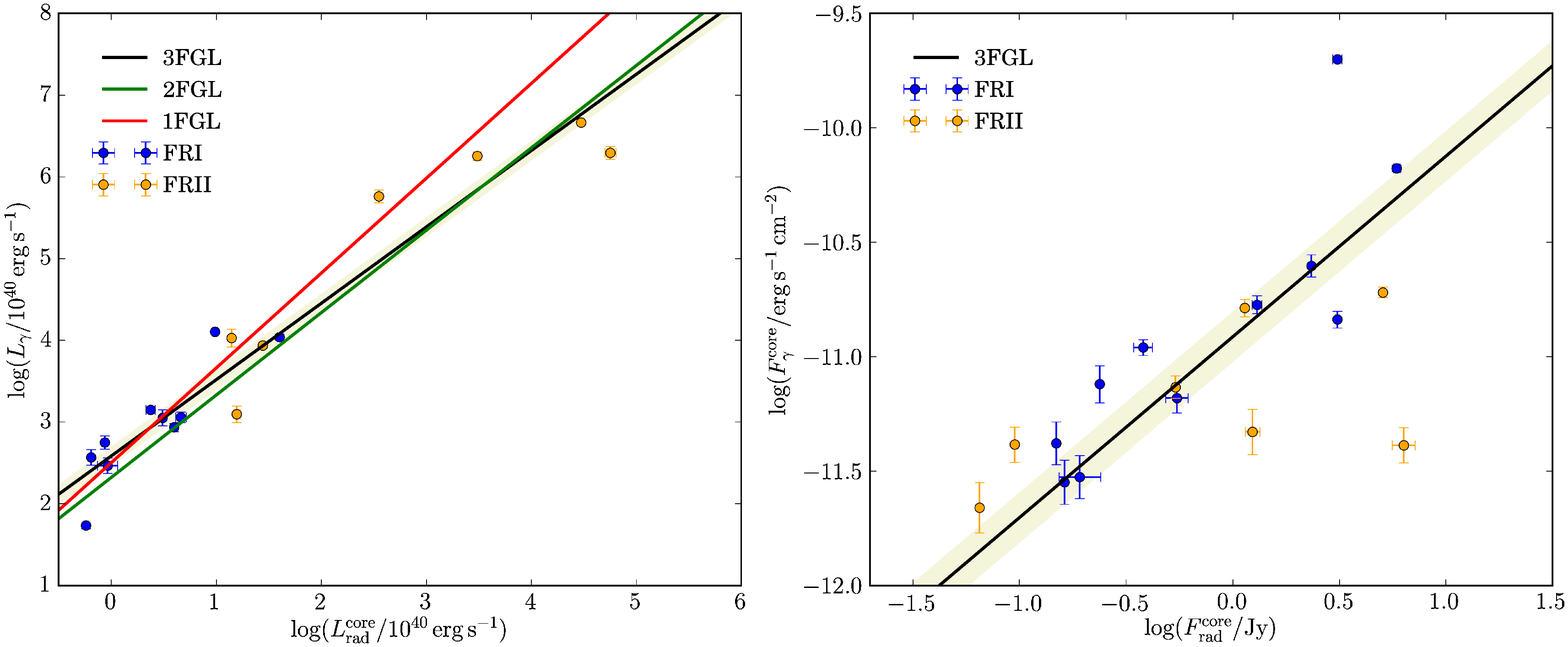}
\caption{Correlation between $\gamma$-ray luminosity/flux above 100 MeV and \textit{core-only} radio luminosity/flux at 5 GHz of RGs.
Legends are the same as Fig.3 in \textit{Extended Data}.
}
\end{centering}
\end{figure}
\begin{deluxetable}{llccccccccc}
\tabletypesize{\scriptsize}
\rotate
\tablecaption{Parameters of radio galaxies in 3FGL.}
\tablewidth{0pt}
\tablehead{
\colhead{Object name} 
& \colhead{3FGL name}
& \colhead{$z$}
& \colhead{$S_{\gamma}$}
& \colhead{$F_\gamma$} 
& \colhead{$\Gamma$}
& \colhead{$F_{\rm rad}^{\rm core}$}
& \colhead{$F_{\rm rad}^{\rm tot}$ }
& \colhead{$\alpha$}
& \colhead{Type}
& \colhead{Ref}\\
&  & & ($10^{-9}\rm\,ph\,cm^{-2}\,s^{-1}$) &  ($10^{-12}\,\ergs\rm\,cm^{-2}$) & & (Jy)  & (Jy) &  & &}
\startdata
NGC 1218	& J0308.6+0408	& 0.029	&$7.10\pm2.13$  & $6.59\pm1.07$  & $2.07\pm0.11$  & $0.548\pm0.07$	&$3.44\pm0.20$	& 0.64  & FRI	 & 1,2
\\
IC 310		& J0316.6+4119	& 0.019	& $5.63\pm2.39$   & $7.58\pm1.55$  &$1.90\pm0.14$  & $0.238\pm0.004$	&$0.238\pm0.004$	& 0.75  & FRI$^{\dagger}$	& 3
\\
NGC 1275	& J0319.8+4130	& 0.018	& $256.51\pm4.84$   & $199.05\pm3.93$	& 2.08  & $3.10\pm0.16$ & $23.3$	& 0.78  & FRI	& 4,5
\\
4C+39.12	& J0334.2+3915	& 0.021	& $5.18\pm2.39$   & $4.18\pm1.00$  &$2.11\pm0.17$   & 0.149	&$0.535\pm0.066$	& 0.39  & FRI$^{\dagger}$    & 5,6
\\
TXS 0348+013	& J0351.1+0128	& 1.120	& $8.09\pm4.29$   & $4.57\pm1.11$  &$2.43\pm0.18$  & $\dots^{c}$	&$0.053\pm0.01$	& $\dots^{c}$  & FRII$^{\ddagger}$    & 5
\\
3C 111		& J0418.5+3813	& 0.049	& $43.17\pm10.06$   & $16.30\pm1.50$  &$2.79\pm0.08$  	& 1.14  &$4.98\pm0.618$	& 0.73  &  FR II  & 5,7
\\
Pictor A	& J0519.2-4542	& 0.035	& $9.28\pm4.93$   & $4.68\pm1.20$  &$2.49\pm0.18$  & $1.24\pm0.10$	&$15.5\pm0.47$	& 1.07   & FR II & 1,8
\\
PKS 0625-35	& J0627.0-3529	& 0.055	& $10.54\pm1.60$    & $16.87\pm1.59$  	&$1.87\pm0.06$	& $1.30\pm0.07$	&$2.25\pm0.09$	& 0.65  & FRI$^{\dagger}$  & 1,9
\\
NGC 2484	& J0758.7+3747	& 0.043	& $4.36\pm2.15$   & $2.82\pm0.70$  &$2.16\pm0.16$	& 0.163	&$1.27\pm0.02$	& 0.07  & FRI  & 1,7
\\
3C 207		& J0840.8+1315	& 0.681   &$14.77\pm4.25$   	& $7.35\pm0.89$  &$2.47\pm0.09$	& $0.539\pm0.003$	&$1.43\pm0.07$	& 0.90  &FRII$^{\ddagger}$  & 1,10
\\
3C 264		& J1145.1+1935	& 0.022	& $2.89\pm1.62$   & $2.97\pm0.72$  &$1.98\pm0.20$	& $0.192\pm0.048$	&$2.43\pm0.01$	& 0.46  & FRI  & 1,2
\\
M87	& J1230.9+1224	& 0.004	& $14.36\pm 2.53$   & $14.52\pm1.27$	&$2.04\pm0.07$	& $3.10\pm0.04$  &$81.70\pm4.28$	& 0.79  &FRI  & 1,11
\\
3C 275.1	& J1244.1+1615	& 0.555	& $8.09\pm4.01$    & $4.13\pm0.80$  &$2.43\pm0.16$	& $0.095$	&$0.973\pm0.133$	&0.96   & FRII$^{\ddagger}$  & 5,7
\\
3C 286		& J1330.5+3023	& 0.850   &$7.97\pm4.95$	& $4.10\pm0.80$	&$2.60\pm0.16$	&$6.34\pm0.82$	&$7.43\pm0.37$	& 0.88  & FRII$^{\ddagger}$  & 1, 5
\\
Cen B		& J1346.6-6027	& 0.013	& $41.88\pm9.60$   & $24.92\pm3.00$	& $2.32\pm0.09$	& $2.34$	&$6.14\pm0.97$	& 0.13  & FRI  & 12,13
\\
3C 303		& J1442.6+5156	&0.141	&$1.65\pm1.06$ 	& $2.19\pm0.63$	& $1.91\pm0.18$	& $0.065$	&$1.04\pm0.14$	& 0.95  &FR II  & 5,7
\\
NGC 6251	& J1630.6+8232	& 0.025	& $21.65\pm1.67$   & $10.95\pm0.91$	& 2.22	  & $0.38\pm0.04$	&$0.637\pm0.064$	& 0.72  & FRI  & 14,15
\\
3C 380		& J1829.6+4844	& 0.692	& $35.22\pm4.00$   & $19.08\pm1.03$	& $2.37\pm0.04$	& $5.07\pm0.105 $	&$7.45\pm0.37$	& 0.71   & FRII$^{\ddagger}$  & 1,16
\\
Cen A	& J1325.4-4301$^{a}$	& 0.002	& $170.47\pm15.46$   &$66.5\pm2.1$	& $2.70\pm0.03$	&$5.88\pm0.21$	& $62.8\pm0.099 $	& 0.70  &FRI  & 8,17
\\
  & J1324.0-4330e$^{b}$	& " "	   &$115.30\pm16.71$   & $51.7\pm4.2 $	& $2.53\pm0.05$	& $\cdots$	& $\cdots$	& " "	& " "
\\
\enddata
\tablecomments{
\noindent
Objects are selected from the Clean Sample of Table 5 from Ackermann et al. (2015) \citep{ackermann2015c}. \\
$S_{\gamma}$ --- $\gamma$-ray photon flux above 100 MeV. \\
$F_{\gamma}$ --- $\gamma$-ray energy flux above 100 MeV.
\\
$\Gamma$ --- Photon index power-law index between 0.1--100 GeV.
\\
$F_{\rm rad}^{\rm core}$, $F_{\rm rad}^{\rm tot}$ --- Core, total radio flux at 5 GHz. The error bars indicate measurement errors.
\\
$\alpha$ --- Radio spectral index.
\\
Object type ---
FRI and FRII: type I and II Fanaroff-Riley galaxy. $^{\dagger\,\ddagger}$ Non-standard \citep{dimauro2014}.
\\
%
References ---
1. Kuhr et al. (1981) \citep{kuhr1981};
2. Kharb et al. (2009) \citep{kharb2009};
3. Kadler et al. (2012) \citep{kadler2012};
4. Taylor et al. (2006) \citep{taylor2006};
5. Gregory \& Condon (1991) \citep{gregory1991};
6. Giovannini et al. (2001) \citep{giovannini2001};
7. Laurent-Muehleisen et al. (1997) \citep{laurent1997};
8. Tingay et al. (2003) \citep{tingay2003};
9. Murphy et al. (2010) \citep{murphy2010};
10. Mullin et al. (2006) \citep{mullin2006};
11. Nagar et al. (2001) \citep{nagar2001}; 
12. Marshall et al. (2005) \citep{marshall2005};
13. Massardi et al. (2008) \citep{massardi2008};
14. Linford et al. (2012) \citep{linford2012};
15. Evans et al. (2005) \citep{evans2005};
16. Mantovani et al. (2009) \citep{mantovani2009};
17. Wright et al. (1994) \citep{wright1994}.
}
\tablenotetext{a,b} { Cen A core and lobes, respectively.}
\tablenotetext{c} { No measurement is given.}
\end{deluxetable}
%
\begin{deluxetable}{lcccccc}
\tabletypesize{\scriptsize}
\tablecaption{Comparison of correlation coefficients}
\tablewidth{0pt}
\tablehead{
\colhead{ }
&\colhead{$\lg$-$\lrad$ }
&\colhead{$p$-value}
&\colhead{$\lg$-$\lrad$($z$) }
&\colhead{$\fg$-$\frad$}
&\colhead{$p$-value}
&\colhead{$\fg$-$\frad$($z$)}
}
\startdata
Total	& $0.863$  & $1.957\times10^{-6}$  & $0.680$  & $0.511$  & $0.026$  & $0.547$
\\  
Core  & $0.938$  & $8.94\times10^{-9}$  & $0.775$  & $0.676$  & $ 0.002$  & $0.752$
\\
\enddata
\tablecomments{
We compare the correlation between $\gamma$-ray and \textit{total}, \textit{core-only} radio luminosities and fluxes.
$p$-values are given in columns next to corresponding Spearman coefficients.
In the last column, partial coefficients are given to exclude the dependence on redshift $z$.
}
\end{deluxetable}


\vskip 0.15in

\small
\noindent
\begin{thebibliography}{}
%
\bibitem{kraushaar1972} Kraushaar, W. L. \textit{et al.} High-energy cosmic gamma-ray observations from the OSO-3 satellite. {\it Astrophys. J.} {\bf 177}, 341-363 (1972).
%
\bibitem{ackermann2015} Ackermann, M. \textit{et al.}. The spectrum of isotropic diffuse gamma-ray emission between 100 MeV and 820 GeV. {\it Astrophys. J.} {\bf 799} 86-110 (2015).
%
\bibitem{ajello2015} Ajello, M. \textit{et al.} The origin of the extragalactic gamma-ray background and implications for dark matter annihilation. {\it Astrophys. J.} {\bf 800}, L27-L34 (2015).
%
\bibitem{fabian2012} Fabian, A. C. Observational evidence of active galactic nuclei feedback. {\it Annu. Rev. Astron. Astrophys.} {\bf 50}, 455-489 (2012).
%
\bibitem{arav2013} Arav, N. \textit{et al.} Quasar outflows and AGN feedback in the extreme UV: HST/COS observations of HE 0238-1904. {\it Mon. Not. R. Astron. Soc.} {\bf 436}, 3286-3305 (2013).
%
\bibitem{cicone2014} Cicone, C. \textit{et al.} Massive molecular outflows and evidence for AGN feedback from CO observations. {\it Astron. Astrophys.} {\bf 562}, 21-46 (2014).
%
\bibitem{tombesi2015} Tombesi, F. \textit{et al.} Wind from the black-hole accretion disk driving a molecular outflow in an active galaxy. {\it Nature} {\bf 519}, 436-438 (2015).
%
\bibitem{zubovas2012} Zubovas, K. \& King, A. Clearing out a galaxy. {\it Astrophys. J.} {\bf 745}, L34-L39 (2012).
%
\bibitem{king2015} King, A. \& Pounds, K. Powerful outflows and feedback from active galactic nuclei. {\it Annu. Rev. Astron. Astrophys.} {\bf 53}, 115-154 (2015).
%
\bibitem{fornasa2015} Fornasa, M. \& Sanchez-Conde, M. A. The nature of the diffuse gamma-ray background. {\it Phys. Reports} {\bf 598}, 1-58 (2015).
%
\bibitem{ackermann2013} Ackermann, M. \textit{et al.} Detection of the characteristic pion-decay signature in supernova remnants. {\it Science} {\bf 339}, 807-811 (2013).
%
\bibitem{caprioli2012} Caprioli, D. Cosmic-ray acceleration in supernova remnants: non-linear theory revised. {\it J. Cosmo. Astrop. Phys} {\bf 07}, 38-56 (2012).
%
\bibitem{ackermann2015b} Ackermann, M. \textit{et al.} Search for early gamma-ray production in supernovae located in a dense circumstellar medium with Fermi LAT. {\it Astrophys. J.} {\bf 807}, 169-184 (2015).
%
\bibitem{caprioli2014} Caprioli, D. \& Spitkovsky, A. Simulatioms of ion acceleration at non-relativistic shocks. I. Acceleration efficiency. {\it Astrophys. J.} {\bf 783}, 91-108 (2014).
%
\bibitem{blandford1987} Blandford, R. \& Eichler, D. Particle acceleration at astrophysical shocks: a theory of cosmic ray origin. {\it Physics Reports} {\bf 154}, 1-75 (1987).
%
\bibitem{chevalier1998} Chevalier, R. A. Sychrotron self-absorption in radio supernovae. {\it Astrophys. J.} {\bf 499}, 810-819 (1998).
%
\bibitem{wang2015} Wang, X. \& Loeb, A. Probing the gaseous halo of galaxies through non-thermal emission from AGN-driven outflows. {\it Mon. Not. R. Astron. Soc.} {\bf 453}, 837-848 (2015).
%
\bibitem{hopkins2007} Hopkins, P. F., Richards, G. T. \& Hernquist, L. An observational determination of the bolometric quasar luminosity function. {\it Astrophys. J.} {\bf 654}, 731-753 (2007).
%
\bibitem{yu2002} Yu, Q. \& Tremaine, S. Observational constraints on growth of massive black holes. {\it Mon. Not. R. Astron. Soc.} {\bf 335}, 965-976 (2002).
%
\bibitem{stecker2007} Stecker, F. W., Malkan, M. A. \& Scully, S. T. Intergalactic photon spectra from the far-IR to the UV Lyman limit for $0<z<6$ and the optical depth of the Universe to high-energy gamma rays. {\it Astrophys. J.} {\bf 658}, 1392-1392 (2007).
%
\bibitem{ackermann2015c} Ackermann, M. \textit{et al.} The third catalog of active galactic nuclei detected by the Fermi Large Area Telescope. {\it Astrophys. J.} {\bf 810}, 14-48 (2015).
%
\bibitem{inoue2011} Inoue, Y. Contribution of gamma-ray-loud radio galaxies' core emissions to the cosmic MeV and GeV gamma-ray background radiation. {\it Astrophys. J.} {\bf 733}, 66-75 (2011).
%
\bibitem{dimauro2014} Di Mauro, M. \textit{et al.} Diffuse $\gamma$-ray emission from misaligned active galactic nuclei. {\it Astrophys. J.} {\bf 780}, 161-175 (2014).
%
\bibitem{ackermann2012} Ackermann, M. \textit{et al.} GeV observations of star-forming galaxies with the Fermi Large Area Telescope. {\it Astrophys. J.} {\bf 755}, 164-187 (2012).
%
\bibitem{su2010} Su, M., Slatyer, T. R., \& Finkbeiner, D. P. Giant gamma-ray bubbles from Fermi-LAT: active galactic nucleus activity or bipolar galactic wind? {\it Astrophys. J.} {\bf 724}, 1044-1082 (2010).
%
\bibitem{crocker2011} Crocker, R. M. \& Aharonian, F. Fermi bubbles: giant, multibillion-year-old reservoirs of Galactic center cosmic rays. {\it Phys. Rev. Lett.} {\bf 106} 101102 (2011).
%
\bibitem{crocker2015} Crocker, R. M. \textit{et al.} A unified model of the Fermi Bubbles, microwave haze, and polarized radio lobes: reverse shocks in the Galactic center's giant outflows. {\it Astrophys. J.} {\bf 808}, 107-136 (2015).
%
\bibitem{cillis2004} Cillis, A. N., Hartman, R. C. \& Bertsch, D. L. Stacking searches for gamma-ray emission above 100 MeV from radio and Seyfert galaxies. {\it Astrophys. J.} {\bf 601}, 142-150 (2004).
%
\setcounter{firstbib}{\value{NAT@ctr}}
\end{thebibliography}

\begin{thebibliography}{}
%
\setcounter{NAT@ctr}{\value{firstbib}}
%
\bibitem{fgq2012} Faucher-Gigu{\`e}re, C.-A. \& Quataert, E. The physics of galactic winds driven by active galactic nuclei. {\it Mon. Not. R. Astron. Soc.} {\bf 425}, 605-622 (2012).
%
\bibitem{guillochon2015} Guillochon, J. \& Loeb, A. The fastest unbound stars in the Universe. {\it Astrophys. J.} {\bf 806}, 124-145 (2015).
%
\bibitem{aharonian2000} Aharonian, F. A. \& Atoyan, A. M. Broad-band diffuse gamma ray emission of the galactic disk. {\it Astron. Astrophys.} {\bf 362}, 937-952 (2000).
%
\bibitem{kelner2006} Kelner, S. R., Aharonian, F. A. \& Bugayov, V. V. Energy specra of gamma rays, electrons, and neutrinos produced at proton-proton interactions in the very high energy regime. {\it Phys. Rev. D} {\bf 74}, 034018 (2006).
%
\bibitem{sturner1997} Sturner, S. J., \textit{et al.} Temporal evolution of nonthermal spectra from supernova remnants. {\it Astrophys. J.} {\bf 490}, 619-632 (1997).
%
\bibitem{carroll1992} Carroll, S. M., Press, W. H. \& Turner, E. L. The Cosmological Constant. {\it Annu. Rev. Astron. Astrophys.} {\bf 30}, 99-542 (1992).
%
\bibitem{abdo2010a} Abdo, A. A., \textit{et al.} Fermi Large Area Telescope observations of misaligned active galactic nuclei. {\it Astrophys. J.} {\bf 720}, 912-922 (2010).
%
\bibitem{abdo2010b} Abdo, A. A., \textit{et al.} Fermi gamma-ray imaging of a radio galaxy. {\it Science} {\bf 328}, 725-729 (2010).
%
\bibitem{akritas1996} Akritas, M. G. \& Bershady, M. A. Linear regression for astronomical data with measurement errors and intrinsic scatter. {\it Astrophys. J.} {\bf 470}, 706-714 (1996).
%
\bibitem{willott2001} Willott, C. J., \textit{et al.} The radio luminosity function from the low-frequency 3CRR, 6CE and 7CRS complete samples. {\it Mon. Not. R. Astron. Soc.} {\bf 322}, 536-552 (2001).
%
\bibitem{lara2004} Lara, L., \textit{et al.} A new sample of large angular size radio galaxies. III. Statistics and evolution of the grown popluation. {\it Astron. Astrophys.} {\bf 421}, 899-911 (2004).
%
\bibitem{venters2009} Venters, T. M., Pavlidou, V. \& Reyes, L. C. The extragalactic background light absorption feature in the blazar component of the extragalactic gamma-ray background. {\it Astrophys. J.} {\bf 703}, 1939-1946 (2009).
%
\bibitem{abdo2010c} Abdo, A. A., \textit{et al.} The first catalog of active galactic nuclei detected by the Fermi Large Area Telescope. {\it Astrophys. J.} {\bf 715}, 429-457 (2010).
%
\bibitem{kuhr1981} Kuhr, H., \textit{et al.} A catalogue of extragalactic radio sources having flux densities greater than 1 Jy at 5GHz. {\it Astron.  Astrophys.} {\bf 45}, 367-430 (1981).
%
\bibitem{kharb2009} Kharb, P., \textit{et al.} Rotation measures across parsec-scale jets of Fanaroff-Riley type I radio galaxies. {\it Astrophys. J.} {\bf 694}, 1485-1497 (2009).
%
\bibitem{kadler2012} Kadler, M., \textit{et al.} The blazar-like radio structure of the TeV source IC 310. {\it Astron. Astrophys.} {\bf 538}, 1-4 (2012).
%
\bibitem{taylor2006} Taylor, G. B., \textit{et al.} Magnetic fields in the centre of the Perseus cluster. {\it Mon. Not. R. Astron. Soc.} {\bf 368}, 1500-1506 (2006).
%
\bibitem{gregory1991} Gregory, P. C. \& Condon, J. J. The 87GB catalog of radio sources covering delta between O and + 75 deg at 4.85 GHz. {\it Astrophys. J. Suppl. Series} {\bf 75}, 1011-1291 (1991).
%
\bibitem{giovannini2001} Giovannini, G., \textit{et al.} VLBI observations of a complete sample of radio galaxies: 10 years later. {\it Astrophys. J.} {\bf 552}, 508-526 (2001).
%
\bibitem{laurent1997} Laurent-Muehleisen, S. A., \textit{et al.} Radio-loud active galaxies in the northern ROSAT All-Sky Survey. I. Radio identifications. {\it Astrophys. J. Suppl. Series} {\bf 122}, 235-247 (1997).
%
\bibitem{tingay2003} Tingay, S. J., \textit{et al.} ATCA monitoring observations of 202 compact radio sources in support of the VSOP AGN Survey. {\it Publ. Astron. Soc. Japan} {\bf 55}, 351-384 (2003).
%
\bibitem{murphy2010} Murphy, T., \textit{et al.} The Australia Telescope 20 GHz Survey: the source catalogue. {\it Mon. Not. R. Astron. Soc.} {\bf 402}, 2403-2423 (2010).
%
\bibitem{mullin2006} Mullin, L. M., Hardcastle, M. J \& Riley, J. M. High-resolution observation of radio sources with $0.6<z\le1.0$. {\it Mon. Not. R. Astron. Soc.} {\bf 372}, 113-135 (2006).
%
\bibitem{nagar2001} Nagar, N. M., Wilson, A. S. \& Falcke, H. Evidence for jet domination of the nuclear radio emission in low-luminosity active galactic nuclei. {\it Astrophys. J.} {\bf 559}, L87-L90 (2001).
%
\bibitem{marshall2005} Marshall, H. L., \textit{et al.} A Chandra survey of quasar jets: first results. {\it Astrophys. J. Suppl. Series} {\bf 156}, 13-33 (2005).
%
\bibitem{massardi2008} Marssardi, M., \textit{et al.} The Australia Telescope 20-GHz (AT20G) Survey: the bright source sample. {\it Mon. Not. R. Astron. Soc.} {\bf 384}, 775-802 (2008).
%
\bibitem{linford2012} Linford, J. D., \textit{et al.} Contemporaneous VLBA 5 GHz observations of Large Area Telescope detected blazars. {\it Astrophys. J.} {\bf 744}, 177-198 (2012).
%
\bibitem{evans2005} Evans, D. A., \textit{et al.} Chandra and XMM-Newton observations of NGC 6251. {\it Mon. Not. R. Astron. Soc.} {\bf 359}, 363-382 (2005).
%
\bibitem{mantovani2009} Mantovani, F., \textit{et al.} Effelsberg 100-m polarimetric observations of a sample of compact steep-spectrum sources. {\it Astron. Astrophys.} {\bf 502}, 61-65 (2009).
%
\bibitem{wright1994} Wright, A. E., \textit{et al.} The Parkes-MIT-NRAO (PMN) surveys. 2: Source catalog for the southern survey (delta greater than -87.5 deg and less than -37 deg). {\it Astrophys. J. Suppl. Series} {\bf 91}, 111-308 (1994).
%
\end{thebibliography}
\end{document}